\documentclass[conference]{IEEEtran}
\IEEEoverridecommandlockouts
\usepackage{times}
\pdfoutput=1
\let\chapter\section
\renewcommand\dbltextfloatsep{3pt}
\renewcommand\dblfloatsep{3pt}
\renewcommand\textfloatsep{3pt}
\usepackage{times}

\usepackage[numbers]{natbib}
\usepackage{multicol}
\usepackage[bookmarks=true]{hyperref}
\usepackage{amsfonts}       
\usepackage{algorithm}
\usepackage{amsmath}
\usepackage{subfigure, graphicx}

\newtheorem{theorem}{Theorem}

\allowdisplaybreaks

\begin{document}
\bstctlcite{IEEEexample:BSTcontrol}
\title{A Separation-Based Design to Data-Driven Control for Large-Scale Partially Observed Systems}

\author
{
Dan Yu$^{1}$, Mohammadhussein Rafieisakhaei$^{2}$ and Suman Chakravorty$^{1}$
\thanks{*This material is based upon work partially supported by NSF under Contract Nos. CNS-1646449 and Science \& Technology Center Grant CCF-0939370, the U.S. Army Research Office under Contract No. W911NF-15-1-0279, and NPRP grant NPRP 8-1531-2-651 from the Qatar National Research Fund, a member of Qatar Foundation, AFOSR contract Dynamic Data Driven Application Systems (DDDAS) contract FA9550-17-1-0068 and NSF NRI project ECCS-1637889.}
\thanks{$^{1}$D. Yu and S. Chakravorty are with the Department of Aerospace Engineering, and $^{2}$M. Rafieisakhaei is with the Department of Electrical and Computer Engineering, Texas A\&M University, College Station, Texas, 77840 USA.
        \{\tt\small yudan198811@hotmail.com, mrafieis, schakrav@tamu.edu\}}%
}

\maketitle

\begin{abstract}
This paper studies the partially observed stochastic optimal control problem for systems with state dynamics governed by Partial Differential Equations (PDEs) that leads to an extremely large problem. First, an open-loop deterministic trajectory optimization problem is solved using a black box simulation model of the dynamical system. Next, a Linear Quadratic Gaussian (LQG) controller is designed for the nominal trajectory-dependent linearized system, which is identified using input-output experimental data consisting of the impulse responses of the optimized nominal system. A computational nonlinear heat example is used to illustrate the performance of the approach. 
\end{abstract}

\IEEEpeerreviewmaketitle
\vspace{-5pt}
\section{Introduction}\label{Section 1}\vspace{-3pt}
In this paper, we consider the stochastic control of partially observed nonlinear dynamical systems that are governed by Partial Differential Equations (PDEs). In particular, we propose a novel data-based approach to the solution of very large Partially Observed Markov Decision Processes (POMDPs) wherein the underlying state space is obtained from the discretization of a PDE; problems whose solution has never been hitherto attempted using approximate MDP-based techniques.  

It is well-known that the global optimal solution for MDPs can be found by solving the Hamilton-Jacobi-Bellman (HJB) equation. The solution techniques can be further divided into model-based and model-free techniques, according to whether the solution methodology uses an analytical model of the system or it uses a black box simulation model or actual experiments. The Reinforcement Learning (RL) techniques \cite{RLHD1, RLHD2} that are based on the Differential Dynamic Programming (DDP) or iLQG approach \cite{RLHD4,RLHD5} have shown the potential for RL algorithms to scale to higher dimensional continuous state and control space problems, such as high dimensional robotic task planning and learning problems.

Fundamentally, rather than solving the derived ``Dynamic Programming" problem as in the majority of the approaches above that requires the optimization of the feedback law, our approach is to directly solve the original stochastic optimization problem in a ``separated open-loop/closed-loop" fashion wherein: 1) we solve an open-loop deterministic optimization problem to obtain an optimal nominal trajectory in a model-free fashion, and then 2) we design a closed-loop controller for the resulting linearized time-varying system around the optimal nominal trajectory, again in a model-free fashion. Nonetheless, the above ``divide and conquer" strategy can be shown to be near-optimal \cite{rafieisakhaei2017nearbelief,rafieisakhaei2017near}.

The primary contributions of the proposed approach are:

1) We specify a detailed set of experiments to accomplish the closed-loop controller design for any unknown nonlinear system, no mater how high dimensional. This series of experiments consists of a sequence of input perturbations to collect the impulse responses of the system, first to find an optimized nominal trajectory, and then to recover the Linear Time-Varying (LTV) system corresponding to the perturbations of the nominal system in order to design an LQG controller.

2) In general, for large-scale systems with partially observed states, the system identification algorithms such as time-varying Eigensystem Realization Algorithm (ERA) \cite{tv_era} automatically construct reduced order model of the LTV system, which results in a reduced order estimator and controller. Therefore, even for large-scale systems, such as partially observed systems with dynamics governed by PDEs, the computation of the feedback policy is computationally tractable. For instance, in the partially observed nonlinear heat control problem considered in this paper, the complexity is reduced by $O(10^5)$ when compared to DDP-based RL techniques.

3) We provide a unification of traditional linear and nonlinear optimal control techniques with Adaptive Dynamic Programming (ADP) \cite{adp} and RL techniques in the context of Stochastic Dynamic Programming problems.
\vspace{-5pt}
\section{Problem Setup}\label{Section 2}\vspace{-3pt}
Consider a discrete-time nonlinear dynamical system:
\begin{align} \label{original system}
x_{k + 1} = f(x_k, u_k, w_k),~~ y_k = h(x_k , v_k),
\end{align}
where $x \in \mathbb{R}^{n_x}$,  $y \in  \mathbb{R}^{n_y}$, $u \in \mathbb{R}^{n_u}$ are the state, measurement and control vectors, respectively, the system and measurement functions, $f(\cdot)$ and $h(\cdot)$, are nonlinear, and $\{w_k, v_k, k\ge 0\}$ are zero-mean, uncorrelated Gaussian white noises with covariances $W$ and $V$, respectively. In considering PDEs, the dynamics are discretized using Finite Difference (FD) or Finite Element (FE), which can lead to a state space problem consisting of, e.g.,  millions of states.

The belief $b(x_k)$ is the conditional distribution of the state $x_k$ given all past data. In this paper, we consider Gaussian beliefs denoted by $b_k := (\mu_k, \Sigma_k)$, where $\mu_k$ and $\Sigma_k$ are the mean and covariance (whose size is $O(n_x^2)$, which for a PDE with large $n_x$ is extremely large ). We denote the belief dynamics by $b_{k + 1} = \tau (b_k, u_k, y_{k + 1})$. 

\textbf{Stochastic Control Problem:} Given $b_0$ and a finite time horizon of $N>0$, for unknown nonlinear $f(\cdot)$ and $h(\cdot)$, find the control policy $\pi = \{\pi_0, \pi_1, \cdots, \pi_{N -1} \}$, where $\pi_k$ is the control policy at time $k$ and $u_k = \pi_k(b_k)$, such that
\begin{align}\label{cost_sto_orig}
J_{\pi} = \mathbb{E}(\sum\limits_{k = 0}^{N - 1} c_k(b_k, u_k) + c_N(b_N)), 
\end{align}
is minimized, where $\{ c_k (\cdot,\cdot) \}_{k = 0}^{N - 1}$ denotes the  immediate cost function, and $c_N(\cdot)$ denotes the terminal cost.
\vspace{-3pt}
\section{Separation-Based Feedback Control Design}\label{Section 3}

Let $ \{{\bar{u}}_k\}_{k=0}^{N-1}, \{{\bar{\mu}}_k\}_{k=0}^{N}, \{{\bar{y}}_k\}_{k=0}^{N}, \{{\bar{b}}_k\}_{k=0}^{N} $ denote the nominal control, state, observation, and belief trajectories of the system, respectively, where given $ {\bar{u}}_{k}={\pi}_{k}({\bar{b}}_k) $, we have:
\begin{align*}
{\bar{\mu}}_{k+1}={f}({\bar{\mu}}_{k},{\bar{u}}_k, 0),
{\bar{y}}_{k}={h}({\bar{\mu}}_{k}, 0),
{\bar{b}}_{k + 1} = {\tau} ({\bar{b}}_k, {\bar{u}}_k, {\bar{y}}_{k + 1}),
\end{align*}
with the initial conditions of $ {\bar{b}}_0 = {b}_0 $, and ${\bar{\mu}}_{0}=\mathbb{E}[{b}_0] $.

The nominal cost and its first order expansion are \cite{rafieisakhaei2017nearbelief,tlqg}:
\begin{align*}
\bar{J}:=&\sum_{k=0}^{N-1}c_k({\bar{b}}_k, {\bar{u}}_k)+c_N({\bar{b}}_N), \nonumber
\\J\approx& \bar{J}\!\! +\!\! \underbrace{\sum_{k=0}^{N-1} ({C}^{{b}}_k({b}_k-{\bar{b}}_k)+ {C}^{{u}}_k({u}_k-{\bar{u}}_k))+ {C}^{{b}}_K({b}_N-{\bar{b}}_N)}_{=:\delta J}.
\end{align*}

\begin{theorem}[Cost Function Linearization Error]\label{theorem:First Order Cost Function Error}  The expected first-order linearization error of the cost function is zero, $\mathbb{E}(\delta J) = 0$. 
\end{theorem}

Theorem \ref{theorem:First Order Cost Function Error} shows that the  first order approximation of the stochastic cost function is dominated by the nominal cost and depends only on the nominal trajectories of the system, independent of the feedback gain. Therefore, the design of the optimal feedback gain can be separated from the design of the optimal nominal trajectory of the system. As a result, the stochastic optimal control problem can be divided into two separate problems: the first is a deterministic problem to design the open-loop optimal control sequence, and hence, the optimal nominal trajectory of the system. The second problem is the design of an optimal linear feedback law to track the nominal trajectory (which is the optimal belief state trajectory unlike typical trajectory optimization based RL methods designed for fully observed problems such as \cite{RLHD1, RLHD2}).

We propose a three-step framework to solve the stochastic feedback control problem as follows.  

\textbf{Step 1. Open-Loop Trajectory Optimization in Belief Space.} 
Solve the open-loop optimization problem given $b_0$:
\begin{align}\label{cost_open}
\{ u_k^*\}_{k = 0}^{N - 1} = &\null  \arg \min_{\{u_k\}} \bar{J} (\{b_k\}_{k=0}^N, \{u_k\}_{k=0}^{N-1}), \nonumber \\
&\null b_{k + 1} = \tau(b_k, u_k, \bar{y}_{k + 1}), 
\end{align}
where the nominal observations $\bar{y}_k$ are generated as follows: $x_{k + 1} = f(x_k, u_k, 0), \bar{y}_k = h(x_k, 0)$ with $x_0 = \mu_0$. Given the nominal observations $\bar{y}_k$, the belief evolution is deterministic and the above is a deterministic optimization problem \cite{platt10}.

The open-loop optimization problem is solved using the gradient descent approach \cite{gradient, sim_opt} utilizing an Ensemble Kalman Filter (EnKF)  \cite{enkf1}. 
Denote  the initial guess of the control sequence by $U^{(0)} = \{u_k^{(0)} \}_{k = 0}^{N - 1}$, and the corresponding belief state estimated using EnKF by $\mathcal{B}^{(0)} = \{b_k^{(0)} \}_{k=0}^N $. The control policy is updated iteratively via
\begin{align}
U^{(n + 1)} = U^{(n)} - \alpha \nabla_U \bar{J}(\mathcal{B}^{(n)}, U^{(n)}),
\end{align}
until a convergence criterion is met, where $U^{(n)} = \{u_k^{(n)} \}_{k = 0}^{N - 1}$ is the control sequence in the $n^{th}$ iteration, $\mathcal{B}^{(n)} = \{b_k^{(n)}\}_{k=0}^N$ denotes the corresponding belief, and $\alpha$ is the step-size parameter. Finlly, denote the nominal belief by $\{ \bar{\mu}_k, \bar{\Sigma}_k\}_{k = 0}^N$.

\textbf{Step 2. Linear Time-Varying System Identification.}
We linearize the system (\ref{original system}) around the nominal mean trajectory $\{\bar{\mu}_k\}$. For simplicity, assume that the control and disturbance enter through same channels and the noise is purely additive: 
\begin{align}\label{perturbation system}
& \delta x_{k+ 1} = A_k \delta x_k + B_k (\delta u_k + w_k), \delta y_k = C_k \delta x_k +  v_k, 
\end{align}
where $\delta x_k = x_k  - \bar{\mu}_k, \delta u_k = u_k - \bar{u}_k, \delta y_k = y_k - h(\bar{\mu}_k, 0)$ describe the state, control and measurement deviations from the nominal trajectory respectively, and 
\begin{align}
& A_k = \frac{\partial f(x, u, w)}{\partial x}|_{\bar{\mu}_k, \bar{u}_k, 0}, B_k = \frac{\partial f(x, u, w)}{\partial u}|_{\bar{\mu}_k, \bar{u}_k, 0},  \nonumber \\
& C_k = \frac{\partial h(x, v)}{\partial x}|_{\bar{\mu}_k, 0}. 
\end{align}

We identify the system (\ref{perturbation system}) using impulse responses of the system via the time-varying ERA \cite{tv_era}. Denote the identified system's deviations by
\begin{align}\label{rom}
&\delta a_{k  +1} = \hat{A}_k \delta a_k + \hat{B}_k (\delta u_k +  w_k), 
\;\delta y_k = \hat{C}_k \delta a_k + v_k, 
\end{align}
where $\delta a_k \in \Re^{n_r}$ denotes the reduced order model (ROM) deviation states, and $n_r << n_x$, thereby automatically providing a compact parametrization of the problem. 

\textbf{Step 3. Closed-Loop Controller Design.} Given system (\ref{rom}), we design the closed-loop controller to  follow the optimal nominal trajectory, which is to minimize the cost function
\begin{align}
J_f = \sum_{k = 0}^{N - 1} (\delta \hat{a}_k' Q_k \delta \hat{a}_k+ \delta u_k' R_k \delta u_k ) + \delta \hat{a}_N' Q_N \delta \hat{a}_N,
\end{align}
where $\delta \hat{a}_k$ denotes the estimates of the deviation state $\delta a_k$, $Q_k, Q_N$ are positive definite, and $R_k$ is positive semi-definite. For the linear system (\ref{rom}), the ``separation principle" of linear control theory can be used \cite{dp_bertsekas}, and the design of the optimal linear stochastic controller can be separated into the decoupled design of a KF and a fully observed optimal LQR controller.

\begin{figure*}[!htbp]
\centering
\subfigure[Comparison of Belief Trajectory]{
\includegraphics[width= 0.23 \textwidth]{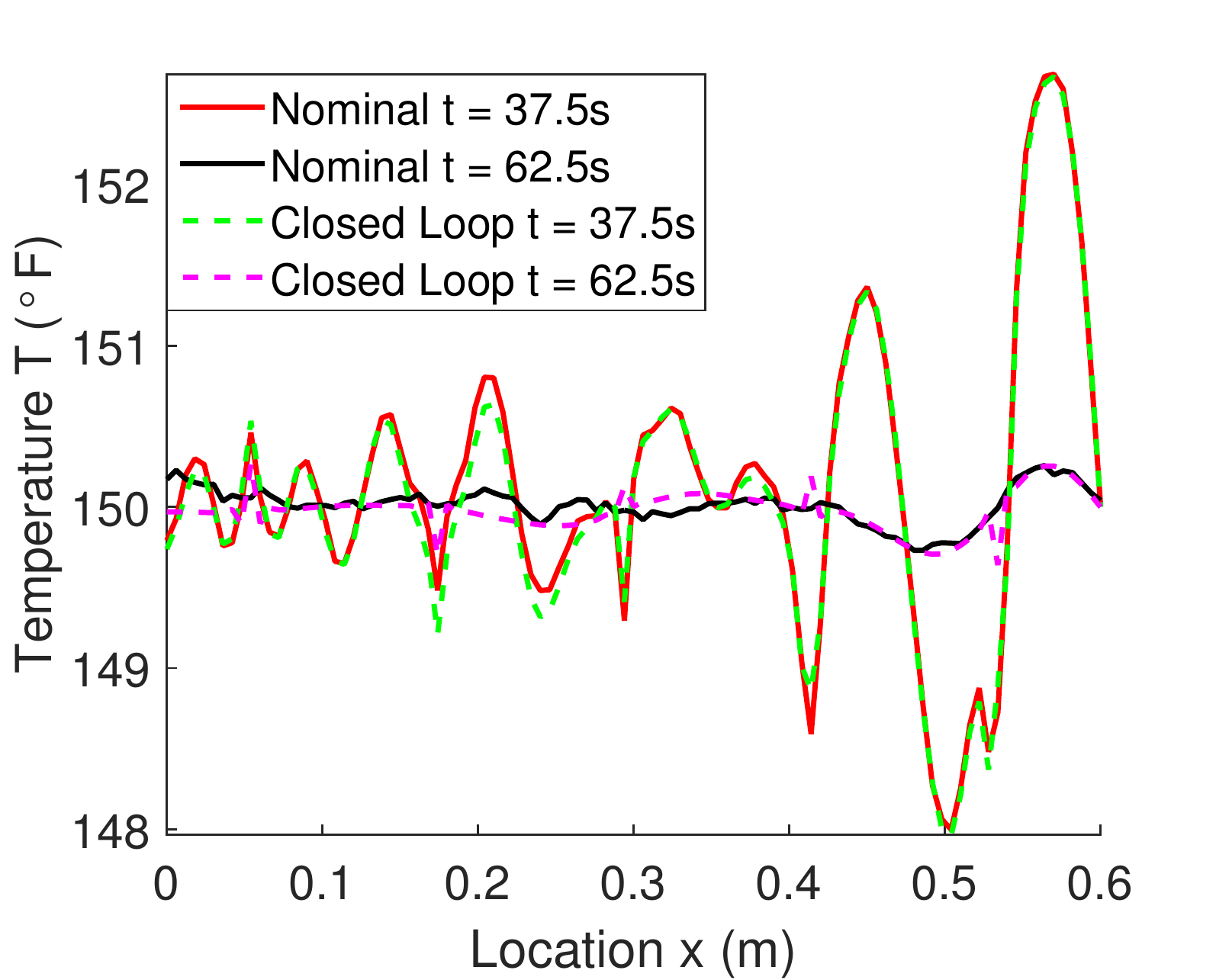}}
\subfigure[Estimation Error at  $x = 0.4L$.]{
\includegraphics[width= 0.23\textwidth]{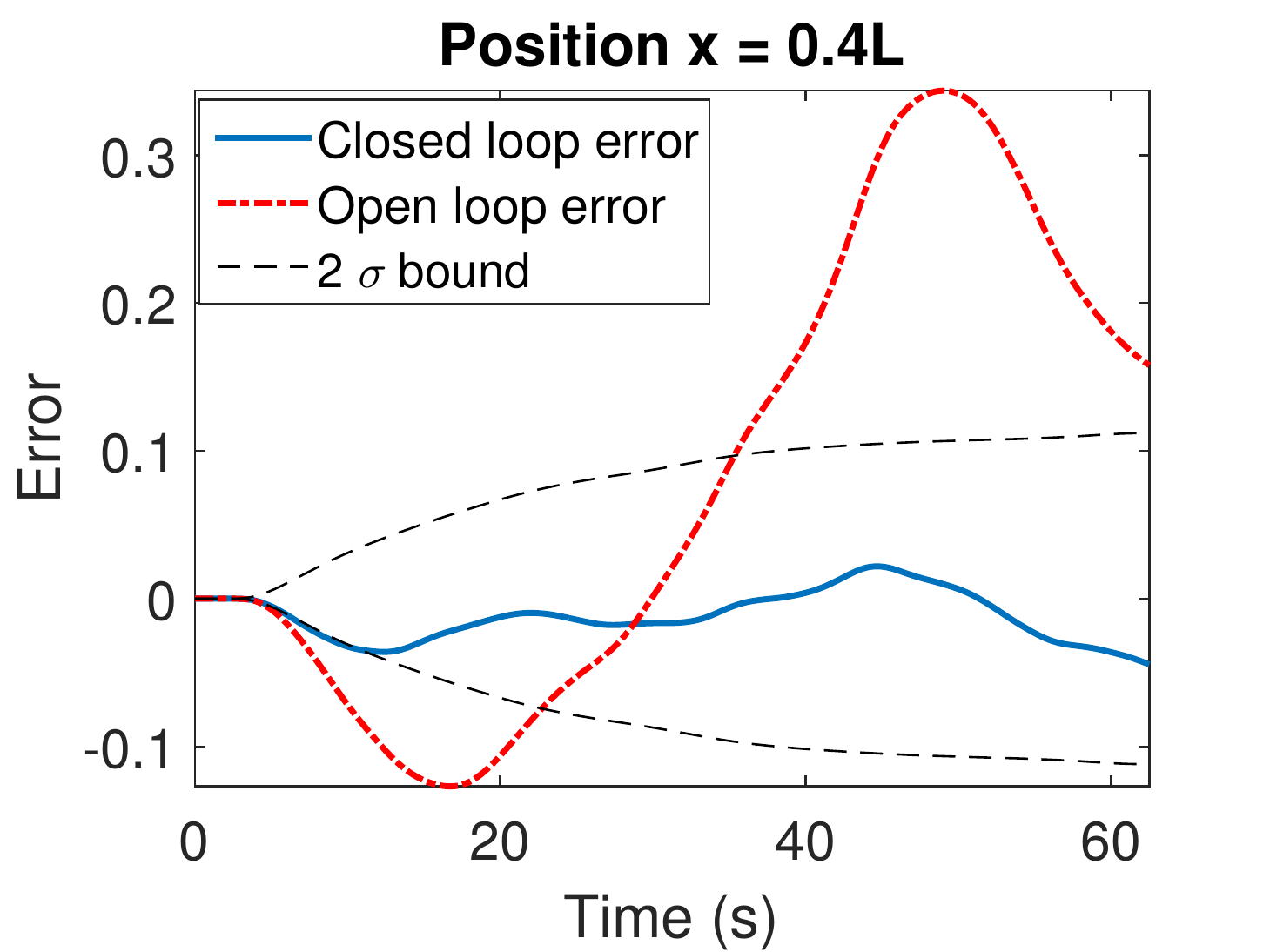}}
\subfigure[Estimation Error at $x = 0.9L$]{
\includegraphics[width= 0.23\textwidth]{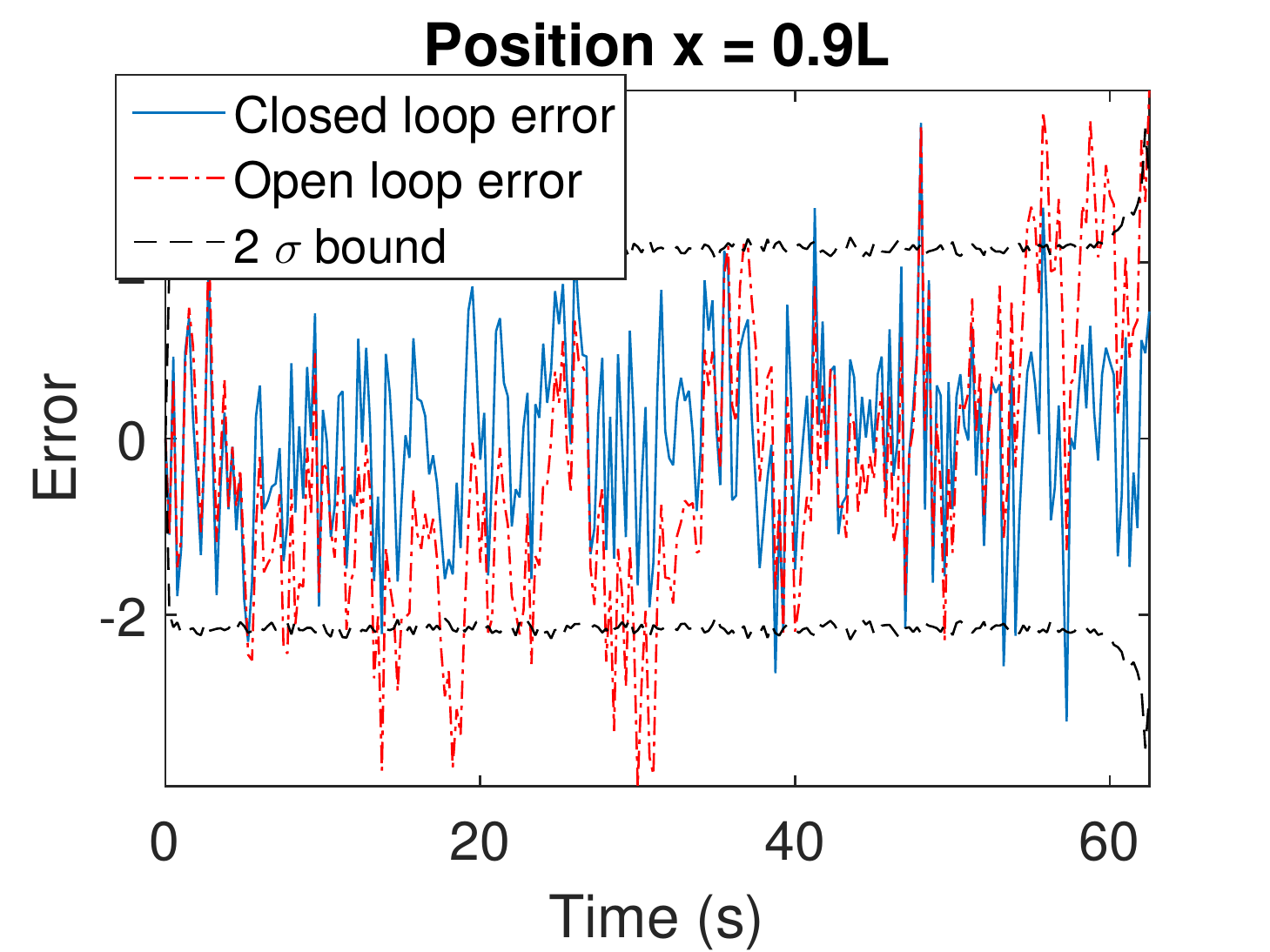}}
\caption{Performance of the Proposed Approach}\label{kf}
\end{figure*}

\begin{figure}[htbp]
\centering
\includegraphics[width= 0.45 \textwidth]{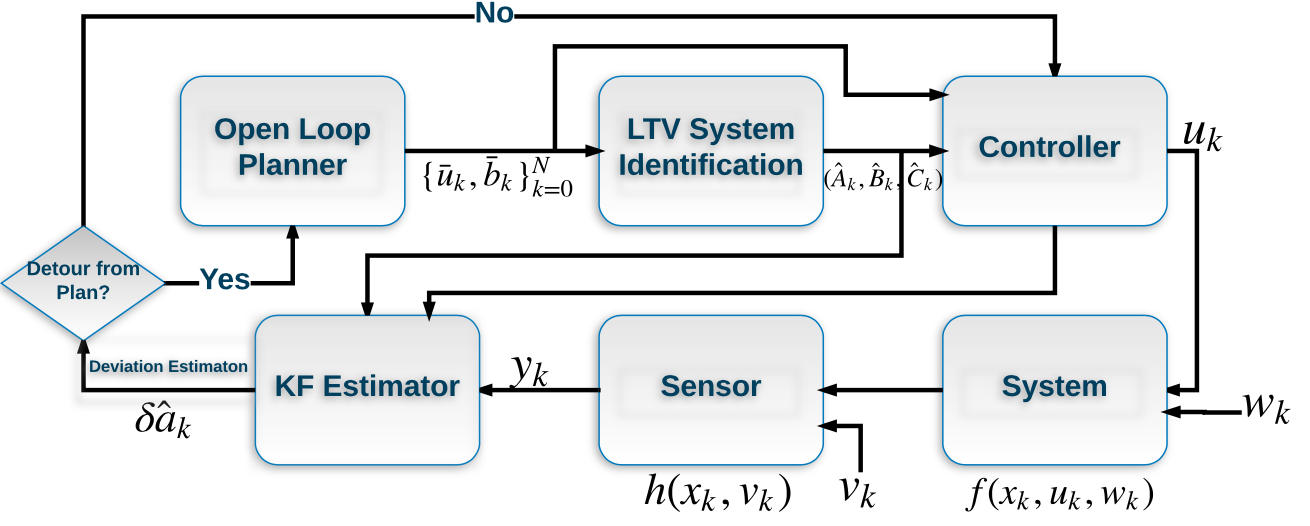}
\caption{Separation-Based Stochastic Feedback Control Algorithm}\label{algo_sc}\vspace{-4pt}
\end{figure}

A flow chart for the Separation-based Nonlinear Stochastic Control Design is shown in Fig. \ref{algo_sc}.
\vspace{-4pt}
\section{Experiments}\label{Section 4}\vspace{-2pt}
We test the method on a one-dimensional nonlinear heat transfer problem. Let $T(x,t)$ be the temperature distribution at location $x$ and time $t$, $K(x, T)$ be the thermal diffusivity, $\eta$ be the convective heat transfer coefficient, $u(t)$ be the external heat sources and $L$ be the length of the slab. The heat transfer PDE along the slab along with its boundary conditions is:
\begin{align}
\frac{\partial T}{\partial t} &= K(x, T) \frac{\partial^2 T}{\partial x^2} - \eta T + u(t),
\\T(x, 0) &= 100 ^{\circ}F,~
\frac{\partial T}{\partial x}|_{x = 0} = 0,~ T(L, t) = 150 ^{\circ}F.
\end{align}

The system is discretized using finite difference method with a 100 equally-spaced grid points. There are five point sources evenly located between $[0.1L, 0.9L]$, where the sensors are placed, as well. The total simulation time is $62.5 s$ with a time-step of $0.25s$. The control objective is to reach the target temperature $T_f\!\! =\!\! (150\! \pm\! 3)^{\circ}F$ for the entire field within $37.5 s$, and keep the temperature at $(150 \pm 3)^{\circ}F$ between $[37.5, 62.5]s$.

\begin{figure}[!htbp]
\centering
\subfigure[Nominal Belief Trajectory]{
\includegraphics[width= 0.23\textwidth]{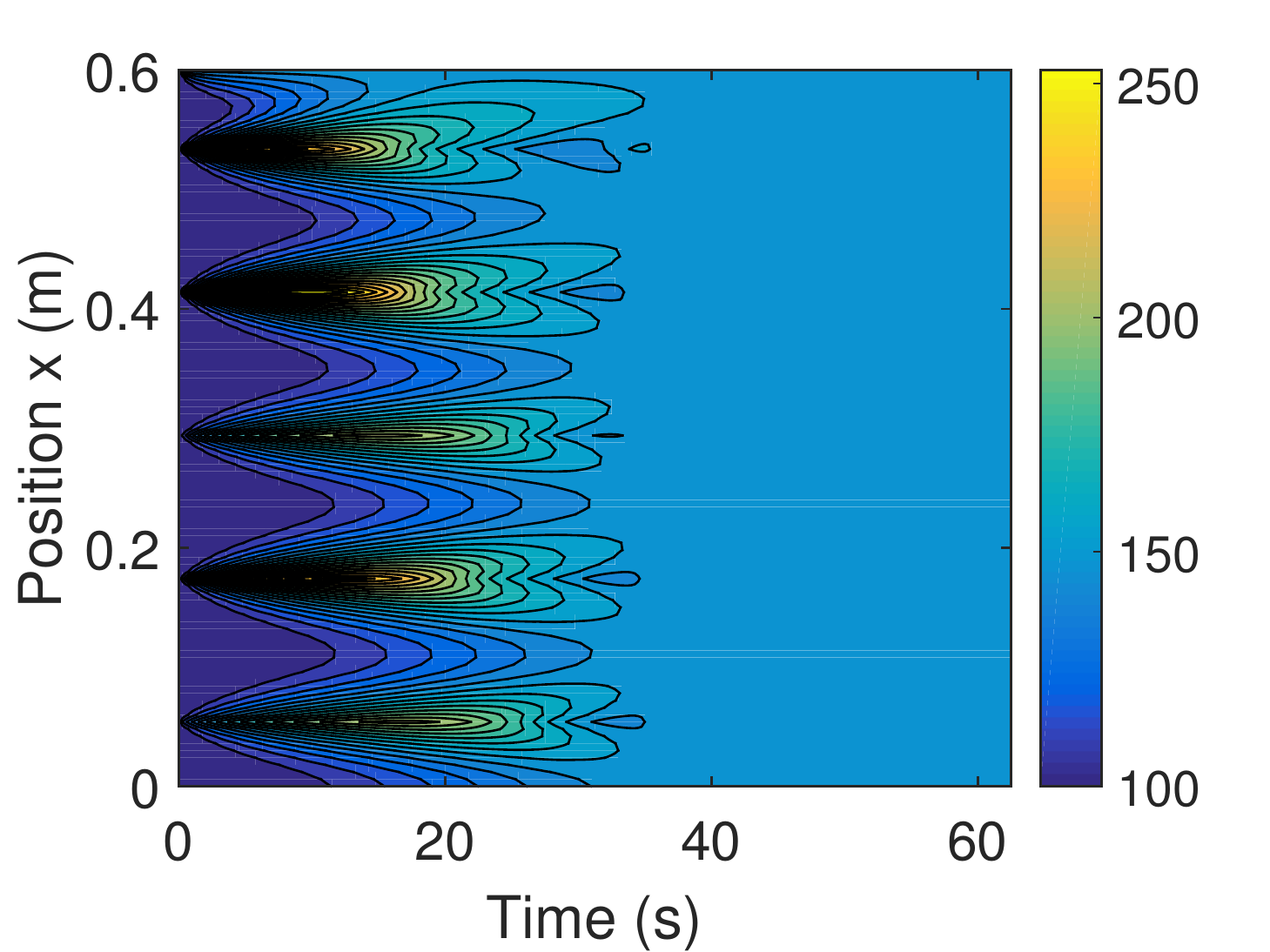}}
\subfigure[Optimal Control]{
\includegraphics[width= 0.23 \textwidth]{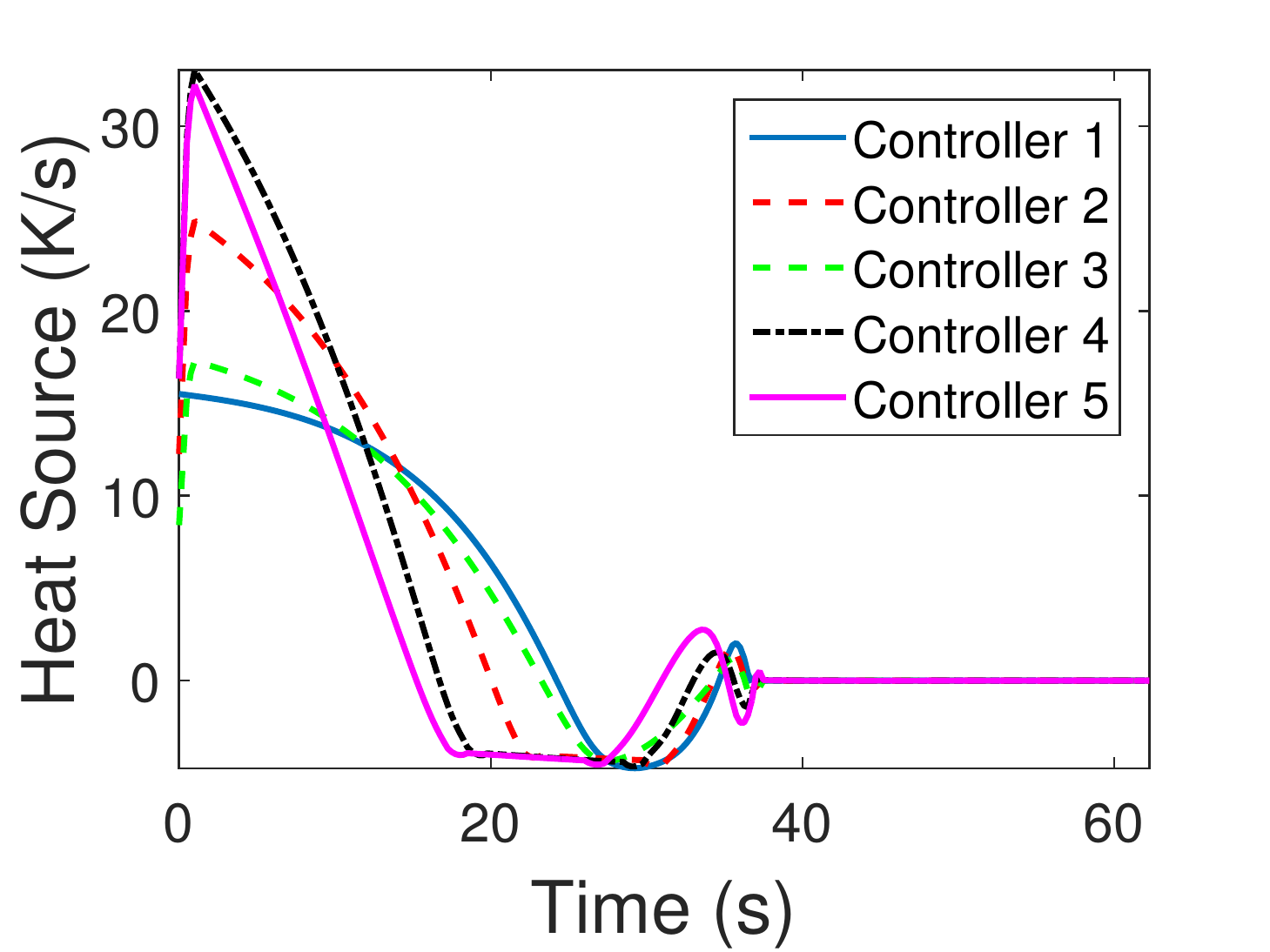}}
\caption{Open Loop Optimization Solution}\label{open}\vspace{-12pt}
\end{figure}

The open-loop optimal nominal (belief mean) trajectory and optimal control are shown in Fig. \ref{open}. For the identified reduced order system, we have $\hat{A}_k \in \Re^{20 \times 20}$. The feedback design decouples into the solution of two 20 x 20 Ricatti equations, one for the controller and one for the Kalman filter. Note that if we were to use an iLQG-based design, the size of the state space would be 10100, and the policy evaluation step would require the solution of a 10100 x 10100 Ricatti equation.\vspace{-15pt}

With the identified linearized system, we design the closed-loop controller. We run 1000 individual simulations with process noise $w_k \sim N(0, I)$ and measurement noise $v_k \sim N(0, I)$. In Fig. \ref{kf}(a), we compare the averaged closed-loop trajectory with the nominal trajectory at time $t = 37.5s, t = 62.5s$. In Figs. \ref{kf}(b)-(c), we randomly choose two positions, and show the errors between the actual trajectory and optimal trajectory with $2 \sigma$ bounds in one simulation. For comparison, the open-loop error is also shown in the figure. 

It is observed that the averaged state estimates over 1000 Monte-Carlo simulations runs are close to the open-loop optimal trajectory, which implies that the control objective to minimize the expected cost function could be achieved using the proposed approach. In this partially observed problem, the computational complexity of designing the online estimator and controller using the identified ROM model is reduced by the order of $O(\frac{ n_x^4}{n_r^2}) = O(10^5)$, and for a general three dimensional problem this reduction could be even more significant.
\vspace{-15pt}
\section{Conclusion}\vspace{-2pt}
In this paper, we proposed a separation-based design of the stochastic optimal control problem for systems with unknown nonlinear dynamics and partially observed states. The open-loop optimization and system identification are efficiently implemented offline using the impulse responses of the system, and an LQG controller based on the ROM is implemented online, which is computationally fast. We showed the performance of the proposed approach on a one-dimensional nonlinear heat transfer problem.

\vspace{-1pt}
\footnotesize{\bibliographystyle{IEEEtran}}
\bibliography{CDC_refs}
\end{document}